\documentclass[twocolumn,pra,aps,superscriptaddress,showpacs]{revtex4-1}
\usepackage{mathptmx}
\usepackage{subfigure}
\usepackage{psfrag,graphicx}
\usepackage{pict2e}
\usepackage{dcolumn}
\usepackage{amsmath,amssymb}
\usepackage{bm}
\usepackage{color}
\usepackage{latexsym}
\usepackage{epstopdf}
\usepackage{color}
\usepackage[english]{babel}
\usepackage{amsfonts}
\usepackage{bm}
\usepackage{natbib}
\usepackage{cancel}
\usepackage{appendix}
\usepackage{bbold}
\usepackage{placeins}
\usepackage{appendix}
\definecolor{mygrey}{gray}{0.35}
\definecolor{myblue}{rgb}{0.2,0.2,0.8}
\definecolor{myzard}{cmyk}{0,0,0.05,0}
\definecolor{mywhite}{rgb}{1,1,1}
\definecolor{mywhite}{rgb}{1,1,1}
\definecolor{myred}{rgb}{1,0.,0.3}
\usepackage[colorlinks=true,citecolor=myblue,linkcolor=myred]{hyperref}
\def\be{\begin{equation}}
\def\ee{\end{equation}}
\def\ba{\begin{align}}
\def\enda{\end{align}}
\def\bi{\begin{itemize}}
\def\ei{\end{itemize}}

\def\adag{{a^{\dag}}}

\begin{document}
\title{Heisenberg scaling with classical long-range correlations}

\author{Samuel Fern\'andez-Lorenzo}
\email{S.Fernandez-Lorenzo@sussex.ac.uk}
\affiliation{Department of Physics and Astronomy, University of Sussex, Falmer, Brighton BN1 9QH, UK}

\author{Jacob A. Dunningham}
\affiliation{Department of Physics and Astronomy, University of Sussex, Falmer, Brighton BN1 9QH, UK}

\author{Diego Porras}
\email{D.Porras@sussex.ac.uk}
\affiliation{Department of Physics and Astronomy, University of Sussex, Falmer, Brighton BN1 9QH, UK}

\date{\today}

\begin{abstract}
The Heisenberg scaling is typically associated with nonclassicality and entanglement. In this work, however, we discuss how classical long-range correlations between lattice sites in many-body systems may lead to a $1/N$ scaling in precision with the number of probes in the context of quantum optical dissipative systems. In particular, we show that networks of coupled single qubit lasers can be mapped onto a classical XY model, and a Heisenberg scaling with the number of sites appears when estimating the amplitude and phase of a weak periodic driving field.
\end{abstract}


\maketitle

\section{Introduction}
Quantum sensing is expected to become one of the key quantum technologies in the short/mid-term, with a wide variety of applications ranging from gravity mapping \cite{abend16} to magnetic detection of single-neuron activity \cite{barry16}. 
In this landscape, quantum resources such as entanglement or nonclassical states of light have been extensively studied as a way to outperform classical resources \cite{giovannetti04,andre04}. In general terms, quantum metrology investigates procedures that accomplish some enhancement in precision, efficiency or simplicity of implementation by means of quantum effects \cite{giovannetti11natphys}. 
For instance, it is now well established that quantum correlations among the initial state of the probes in Ramsey interferometry may surpass the so-called \textit{standard quantum limit} or \textit{shot-noise limit} \cite{giovannetti04}. In this limit, the precision in parameter estimation scales as $1/\sqrt{N}$, where N is the resource count (number of probes in our case). Quantum effects may give rise to an increase in precision to reach the so-called \textit{Heisenberg limit}, which scales as $1/N$.
Frequently, however, these potential benefits are hindered by the effect of noise and decoherence over delicate quantum states \cite{ono10,guta12natcomm}. For example, the incoherent loss of a photon in a NOON state, well-known in optical interferometry for leading to a Heisenberg scaling, turns it into a useless mixed state \cite{giovannetti11natphys}.  \

\begin{figure}[h!]
  \includegraphics[width=0.45\textwidth]{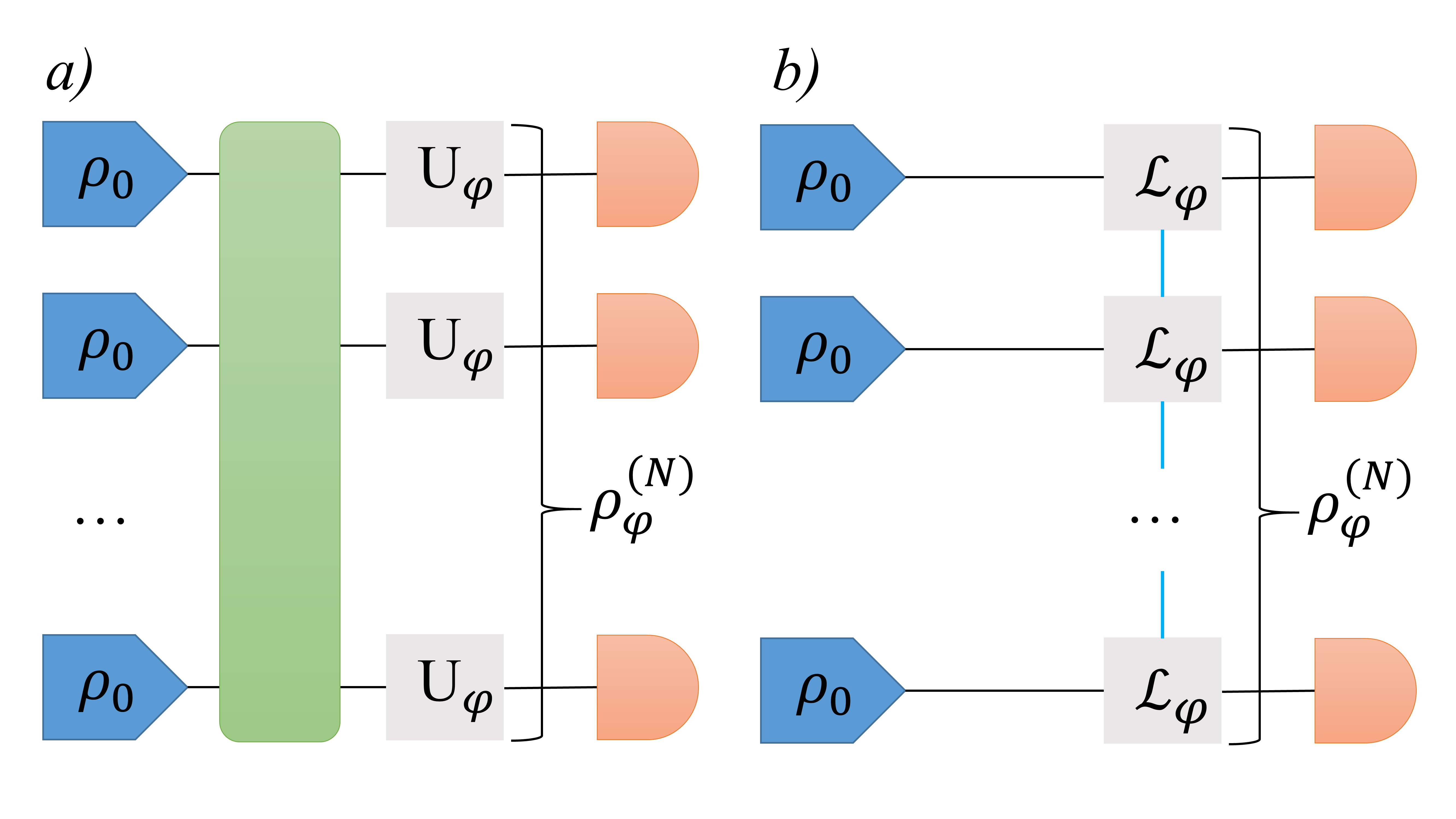}
  \caption{Comparison with Ramsey interferometry \cite{giovannetti11natphys}. $\rho_0$ represents the initial state of N probes and local detections are performed at the end (orange semicircles). a) Entanglement among the probes (green box)	is generated before they are fed into a unitary channel $U_{\varphi}$ that leads to a joint state $\rho^N_{\varphi}$. b) Initial probe states evolve under a Markovian channel $\mathcal{L}_{\varphi}$ with first-neighbor interactions among them (blue lines). The state preparation and the interaction with the probes occur simultaneously.}
  \label{fig:fig1} 
\end{figure}

In the last years, different protocols were conceived to produce robust sensing schemes, such as quantum illumination \cite{lloyd08,lopaeva13,sanz17,Zhuang17} or quantum error correction \cite{kessler14,arrad14}. Ideally, one would like to combine the enhancement given by the Heisenberg scaling with the robustness of classical states. On the one hand, although dissipation is typically considered as an obstacle, it may be turned into an asset to engineer advantageous states for quantum metrology. Useful symmetry properties and criticality exhibited by dissipative phase transitions have been proposed as useful resources for sensing purposes \cite{Lorenzo17,raghunandan17}. This approach has the advantage that no initial state preparation is required and furthermore, the steady state may be naturally robust against noise, which is normally the key limiting factor in other schemes.
On the other hand, one could exploit the correlations naturally developed in many-body systems as an alternative to the initial preparation of quantum correlations in Ramsey interferometry. In particular, lattice systems with local (nearest-neighbors) interactions are now within the state-of-the-art techniques, which enables the study of a rich variety of dissipative phase states and transitions \cite{rota17,Hartmann08}. The potential benefits of local interactions and quantum phase transitions in closed systems haven been already considered \cite{Zanardi08,skotiniotis15}, along with nonlinear estimation strategies in systems like BECs \cite{roy08,choi08,boixo07,maldonado09}. In Fig.\ref{fig:fig1}, all these ideas are schematically compared with the canonical Ramsey interferometer. In this work the target parameter is incorporated in a linear Hamiltonian and it is based on \textit{nearest-neighbor} interactions between sites so that the resources (number of probes) scale with $N$. \

This work presents the following results. (i) We introduce a specific dissipative model of $N$ single qubit lasers with an effective dissipative-mediated coupling in first-neighbors. (ii) The steady state of this model is shown to be formally equivalent to a thermal state of the classical XY model subjected to an external field. (iii) Analytical expressions of optimal observables for estimating the amplitude and phase of a weak periodic driving as well as the corresponding Fisher information are presented. A Heisenberg scaling with the number of lattice sites is manifested as a result of \textit{classical} long-range correlations in the lattice. These long-range correlations are naturally developed, under the right conditions, by the dissipative dynamics with short-range interactions. The type of short-range interactions employed are typically present in networks of quantum optical systems such as superconducting circuits, cavity QED and trapped ions, on which our work is focused. As a result, even though the resources scale with $N$, one yet may achieve a quantum Fisher information scaling as $N^2$, which is compatible with a notion of resource counting and Heisenberg limit based on the Margolus-Levitin bound \cite{zwierz10,zwierz12}.
\section{Lattice of single-qubit lasers}
We shall study a chain of $N$ identical coupled single-qubit lasers. This system is a generalization of our previous scheme in \cite{Lorenzo17}. 
Every single-qubit laser consists of a bosonic mode $a_j$ coupled by a Jaynes-Cummings interaction to a two-level system (qubit), with levels $|g\rangle$ and $|e\rangle$, subjected to incoherent pumping of the qubit and losses of the bosonic mode with rates $\gamma$ and $\kappa$, respectively. These dissipative processes are well-described though appropriate master equations \cite{BreuerPet}, for which the following notation for Lindbald super-operators (dissipators) will be employed,
\begin{equation}
\mathcal{L}_{\{O , \Gamma\}}(\rho)= \Gamma (2 O\rho O^{\dag}-O^{\dag}O\rho - \rho O^{\dag}O).
\end{equation}
Each mode is additionally fed with a weak coherent periodic driving field whose amplitude $|\epsilon|$ and phase $\phi$ are aimed to be estimated. The qubit and the driving frequencies are in resonance with the bosonic modes.

\begin{figure}[h!]
  \includegraphics[width=0.5\textwidth]{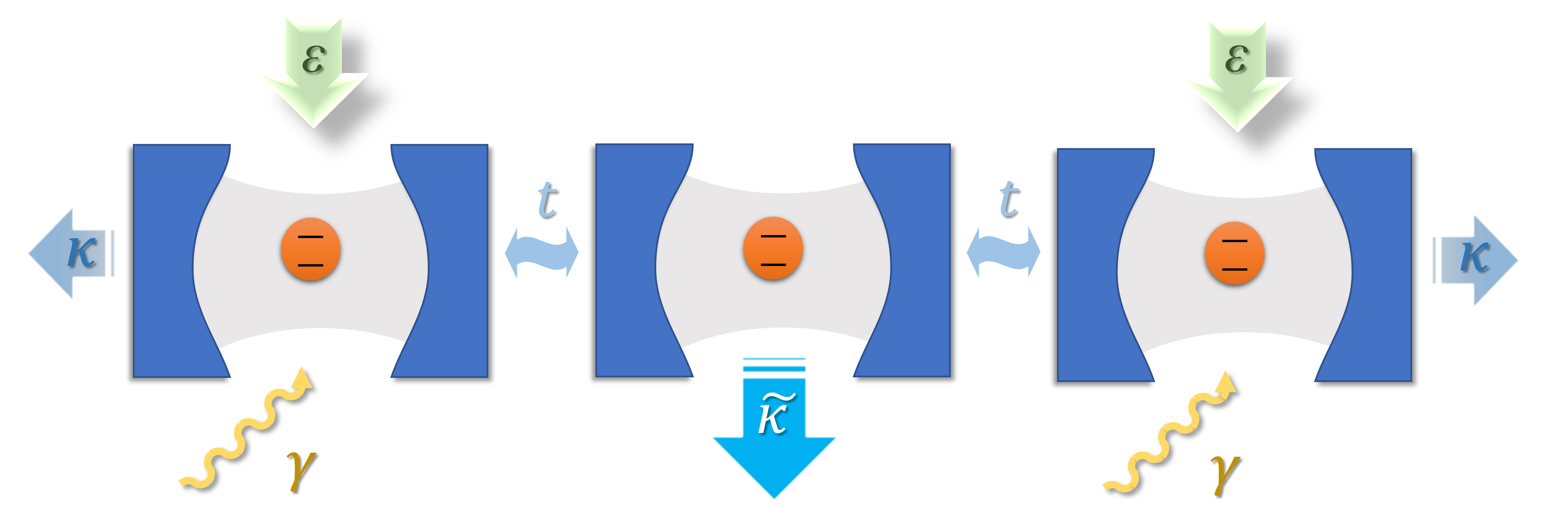}
  \caption{General scheme of the dissipative-mediated coupling. Neighboring single qubit lasers are coupled through a coherent hopping term to a fast decaying mode at rate $\tilde{\kappa}$, which is adiabatically eliminated. Each single qubit laser is subjected to incoherent qubit pumping at rate $\gamma$, mode losses at rate $\kappa$ and a periodic driving field $\epsilon$.}
  \label{fig:fig2} 
\end{figure}

We are interested in implementing an incoherent coupling of each qubit laser with its neighbors, which will induce classical correlations among them. Dissipative couplings appear naturally through evanescent modes in arrays of coupled macroscopic lasers \cite{Oliva01,eckhouse08,nixon11}. However in microscopic systems of single-mode cavity arrays \cite{Hartmann08} or superconducting circuits \cite{houck12natphys}, bosonic modes are coupled by coherent photon tunneling terms. To get a dissipative coupling from these coherent terms, we assume that the cavities are coupled by intermediate auxiliary modes $b_{k}$ with a fast photon decay rate, $\tilde{\kappa}$ (see Fig. \ref{fig:fig2}). The coherent hopping is given by the Hamiltonian term,
\begin{equation}\label{hoppingTerm}
H^{\rm hop}=-t\sum_{\langle k,j\rangle}(a^{\dag}_j b_k+b^{\dag}_k a_{j+1}+h.c.),
\end{equation}
with $t$ being the photon tunneling amplitude. The adiabatic elimination of these auxiliary modes results in an effective dissipative interaction. This can be shown by calculating the Heisenberg equations for $b_k$, yielding
\begin{equation}
\dot{b_k}=-it(a_j+a_{j+1})-\tilde{\kappa}b_k,
\end{equation}
where $a_j$ and $a_{j+1}$ are the neighboring modes. In the case that $b_k$ is a fast decaying mode, i.e., $\tilde{\kappa}\gg 1$, one may adiabatically eliminate it by taking $\dot{b_k}\approx 0$ and using its steady-state solution,
\begin{equation}
b_k=-\frac{it}{\tilde{\kappa}}(a_j+a_{j+1}). \label{interm}
\end{equation}
The substitution of Eq.\eqref{interm} in the complete dynamics will result in the effective elimination of the direct hopping \eqref{hoppingTerm}, whereas the dissipator of the intermediate mode originates an effective dissipative-mediated coupling given by,
\begin{equation}
\mathcal{L}_{\{b_k , \tilde{\kappa}\}}\rightarrow\mathcal{L}_{\{a_j+a_{j+1} , \frac{t^2}{\tilde{\kappa}}\}}.
\end{equation}
 In an interaction picture rotating at the mode frequency and performing such adiabatic elimination, the whole dynamics is described by the following master equation for the system density matrix $\rho$,
\begin{equation}
\dot{\rho}=-i [H,\rho]+\sum^N_{j}\left(\mathcal{L}_{\{\sigma_j^+ , \gamma\}} +\mathcal{L}_{\{a_j , \kappa\}}+\mathcal{L}_{\{a_j+a_{j+1} , \frac{t^2}{\tilde{\kappa}}\}}\right)(\rho)
\label{Liouvillian}
\end{equation}
where the Hamiltonian is given by,
\begin{eqnarray}
H &=& \sum^N_{j} H_j^{\rm JC}+\sum^N_{j}H^{\rm d}_j , \label{Hamiltonian} \nonumber
\\
H_j^{\rm JC} &=& g(\sigma_j^+ a_j+ \adag_j\sigma_j^-), \quad H_j^{\rm d}= \epsilon^* a_j  + \epsilon a_j^{\dag},
\end{eqnarray}
and $\epsilon=|\epsilon|e^{i\phi}$.
Note that the last dissipator in Eq.\eqref{Liouvillian} represents the effective dissipative-mediated coupling in first-neighbors. A mean field calculation of \eqref{Liouvillian} predicts a dissipative phase transition to a lasing phase when the renormalized pumping parameter 
\begin{equation}
\tilde{C}_{\rm p}=\frac{C_{\rm p}}{(1+3(t/\kappa)^2)}
\end{equation}
satisfies $\tilde{C}_{\rm p}>1$ ($C_{\rm p}\equiv g^2/(\kappa\gamma)$) (see appendix \ref{App:AppendixA}).

For sensing purposes, the single qubit laser will be prepared to work in a regime of large number of bosons \cite{Lorenzo17}. This can be accomplished in a strong pumping regime of the two-level systems, \textit{i.e.}, $\gamma\gg g,\kappa,|\epsilon|$, in which the qubits can be adiabatically eliminated \cite{Mandel95}. This leads to the following effective quartic master equation for the bosonic mode (see appendix \ref{App:AppendixB} for details), 
\begin{align} 
	\dot{\rho}_f&=-i\sum_j^N[\epsilon^* a_j  + \epsilon a_j^{\dag},\rho_f]+\sum_j^N\mathcal{L}_{\{a_j+a_{j+1} , D\}}(\rho_f)  \label{adiabatic}  \\ 
&+\sum^N_{j}\left(\mathcal{L}_{\{a^{\dag}_j, A\}}+\mathcal{L}_{\{a_j a^{\dag}_j, B\}}- \mathcal{L}_{\{(a^{\dag}_j)^2, B\}}+\mathcal{L}_{\{a_j, C-\frac{t^2}{\kappa}\}} \right)(\rho_f). \nonumber 
\end{align}
We have introduced the coefficients 
$A=g^2/\gamma$, 
$B=2g^4/\gamma^3$, 
$C=\kappa+D$,
$D=t^2/\tilde{\kappa}$, and $\rho_f = {\rm Tr}_{\rm qubit}{\{\mathcal{L}(\rho)\}}$ is the reduced density matrix of the bosonic field.
Equation \eqref{adiabatic} is valid below the critical point, $\tilde{C}_{\rm p}<1$, and slightly above it, ${C}_{\rm p}\gtrsim 1$.
%

\section{Semi-classical limit}
Equation \eqref{adiabatic} can be more conveniently expressed as an equation in phase space. Concretely, we shall use the \textit{ Glauber-Sudarshan P} representation \cite{Mandel95} of the effective master equation, defined as 
\begin{equation}
\rho(t)=\int d^2\alpha P(\alpha,\alpha^*,t) |\alpha\rangle\langle \alpha |
\end{equation}
where $|\alpha\rangle$ is the coherent state $|\alpha\rangle=\exp{(\alpha a^{\dag}-\alpha^* a)}|0\rangle$. The function $P(\alpha,\alpha^*)$ is a quasi-probability distribution over $|\alpha\rangle\langle \alpha |$, with the normalization condition $\int d^2\alpha P(\alpha,\alpha^*,t) = 1$ and expectation values given by $\langle (\adag)^p a^q \rangle=\int d^2\alpha (\alpha^*)^p \alpha^q P(\alpha,\alpha^*)$. The conversion between the operator master equation \eqref{adiabatic} and its representation in phase space can be carried out thanks to the following equivalences
\begin{align}
	a|\alpha\rangle\langle \alpha |&=\alpha |\alpha\rangle\langle \alpha | \\ 
	|\alpha\rangle\langle \alpha | \adag &=\alpha^* |\alpha\rangle\langle \alpha | \\
	\adag |\alpha\rangle\langle \alpha | &=\left( \frac{\partial}{\partial\alpha} +\alpha^* \right)  |\alpha\rangle\langle \alpha | \label{alphaderiv}\\ 
	|\alpha\rangle\langle \alpha |a &=\left( \frac{\partial}{\partial\alpha^*} +\alpha \right)  |\alpha\rangle\langle \alpha | . \label{alphaderiv2}
\end{align}

In a regime of large number of bosons $|\alpha|^2\gg 1$, the substitution of this representation leads to an equation of motion for $P(\alpha,\alpha^*,t)$ (see Appendix \ref{App:AppendixC} for derivation) with the form of the well-known Fokker-Planck equation \cite{Risken84},
\begin{align} \label{PFokker1}
\frac{\partial P}{\partial t}&=+2A\sum_j\frac{\partial^2P}{\partial \alpha_j\partial\alpha_j^*} \\
 &-\sum_{\langle j,k\rangle}\frac{\partial}{\partial \alpha_j} [(A-C-B|\alpha_j|^2)\alpha_j  -D\alpha_k-\epsilon'] P +c.c. , \nonumber
\end{align}
where $\epsilon'\equiv i\epsilon$ and $\langle j,k\rangle$ stands for first neighbors. Equation \eqref{PFokker1} presents the adequate structure so that the steady state may be analytically integrated using a certain detailed balance condition (see App. \ref{App:AppendixC}). In polar coordinates $\alpha_j=r_je^{i\theta_j}$, the steady state reads as follows,
\begin{equation}\label{steady1}
\begin{split} 
P(\vec{r},\vec{\theta})= \frac{1}{Z} \exp  & \left( \sum_j( \mu r_j^2 -\lambda r_j^4-2\nu r_j\sin{(\theta_j-\phi)} )- \right. \\
  &\quad \left. {} -\sum_{\langle j,k\rangle}2\varsigma r_jr_k \cos{(\theta_j-\theta_k)}  \vphantom{}\right) ,
\end{split}
\end{equation}
where we use the notation  $\vec{r}=(r_1,r_2,\cdots,r_N)$ and $ \vec{\theta}=(\theta_1,\theta_2,\cdots,\theta_N)$. We have also introduced the parameters 
$\lambda = B/2A$,
$\mu=(A-C)/A$,
$\nu=|\epsilon|/A$
 and $\varsigma=D/A$, and $Z$ is a normalization constant. 
%
%

The radial components $r_j$ are essentially associated to the number of bosons in each cavity $r_j^2\approx n_j$. As the input signals $\epsilon_j$ are assumed to be weak, their major influence will be on the angular dynamics, while the radial components $r_j$ will be settled on their steady-state values $r_j\approx r_0$. In this case, the dynamics of Eq. \eqref{PFokker1} will be dominated by the angular components for laser operation sufficiently far above threshold, and one can derive an effective equation for the angular variables $\vec{\theta}$. This can be done by assuming a $P$ function of the form 
\begin{equation}
P(\mathbf{r},\mathbf{\theta})=R(r_1)R(r_2)\cdots R(r_N)P'(\mathbf{\theta}),
\end{equation}
where each $R(r_j)$ is a Gaussian distribution properly normalized around $r_0$. The resulting equation reads (see App. \eqref{App:AppendixC}),
\begin{align} \label{PFokker2}
\frac{\partial P'}{\partial t}&=+\frac{A}{2n_0}\sum_j\frac{\partial^2P'}{\partial \theta_j^2}  \\
&+\sum_{\langle j,k\rangle}\frac{\partial}{\partial \theta_j}\left((D\sin(\theta_k-\theta_j)+\frac{|\epsilon|}{\sqrt{n_0}}\cos(\theta_j-\phi))P'\right) \nonumber,
\end{align}
in which $n_0=r_0^2$ stands for the steady average number of bosons \textit{per site}. Equation \eqref{PFokker2} can be related to the first-neighbors stochastic Kuramoto model of $N$ identical oscillators \cite{Acebron05,Rodrigues16}. The Kuramoto model is paradigmatic in the study of synchronization, and it has gained renewed attention in the context of complex \cite{Rodrigues16,Strogatz01,Arenas06} and neural networks \cite{Cumin07}. The steady state solution to Eq.\eqref{PFokker2} can be obtained by imposing $r_j=r_0$ in \eqref{steady1} and tracing over the radial part,
\begin{equation}\label{steady2}
\begin{split} 
P'(\vec{\theta})= \frac{1}{Z} \exp   \left(  -\sum_{\langle j,k\rangle}2\varsigma  n_0 \cos{(\theta_j-\theta_k)}- \right. \\
  \quad \left. {} -\sum_j 2\nu \sqrt{n_0}\sin{(\theta_j-\phi)}  \vphantom{}\right) .
\end{split}
\end{equation}

We identify the steady state \eqref{steady2} as formally equivalent to a thermal state of an antiferromagnetic classical XY model in the presence of an external field, with an effective temperature 
\begin{equation}
\beta_{\rm eff}=\frac{n_0}{(C_{\rm p}\kappa)}.
\end{equation}
Our setup \eqref{Liouvillian} is thus revealed as an alternative to simulate the XY model, which has been recently implemented in various platforms \cite{struck11,berloff16, nixon13,takeda17,tamate16}.
%
This has been proved to be particularly fruitful in the study of geometric frustration \cite{struck11,nixon13}. 
In the context of machine learning, the XY model has also been suggested as an alternative to Markov chain Montecarlo methods in order to speed up the computationally time-consuming Boltzmann sampling \cite{takeda17,tamate16,ackley85,zemel95,baldi90}. \\

%
%

\section{Quantum Fisher information and Heisenberg scaling}
The maximal resolution that can be achieved by means of the lattice qubit laser for estimating the amplitude $|\epsilon|$ and phase $\phi$ can be systematically assessed in terms of the quantum Fisher information (QFI), $F_Q$ \cite{Demkow14}. This theory sets an ultimate lower bound on the resolution attainable when estimating certain parameter $\varphi$ encoded in a density matrix $\rho_\varphi$ through the well-known quantum \textit{Cramer-Rao bound},
\begin{equation}
\Delta^2\varphi\geq \frac{1}{F_Q[\rho_\varphi]}.
\end{equation}
An observable that saturates this bound is said to be optimal. The so-called \textit{symmetric logarithmic derivative} (SLD), $L_\varphi$, defined through the operator equation  
\begin{equation}
\partial_\varphi \rho_\varphi=\frac{1}{2}(\rho_\varphi L_\varphi+L_\varphi \rho_\varphi),
\end{equation}
gives us such optimal observable \cite{Braunstein96}. The QFI can then be obtained as $F_Q[\rho_\varphi]=Tr\left\{\rho_\varphi L_\varphi^2\right\}$. 
For the single-qubit laser, the optimal observables for estimating  $|\epsilon|$ and $\phi$ in the steady state are the field quadratures  
\begin{align}
 \hat{P}_\phi&= i(ae^{-i\phi}-a^{\dag}e^{i\phi}) \\
\hat{X}_\phi &= (a e^{-i\phi}+a^{\dag}e^{i\phi})
\end{align}
respectively ($([\hat{X}_\phi,\hat{P}_\phi]=2i)$)\cite{Lorenzo17}. \\

We shall first focus on the amplitude estimation for a given phase. It is natural to suggest that the linear combination
\begin{equation}
\hat{P}^\phi_{\rm sum}=\sum_j\hat{P}^\phi_j
\end{equation}
may be the optimal observable for the lattice qubit laser, at least for weak couplings $t$ between sites. By using such linear ansatz for the SLD, it can be shown (see App. \ref{App:AppendixE}) that this assumption is correct as long as $\varsigma\ll 1$, a condition easily satisfied in our setup. 
The analytical expression of $\langle\hat{P}^\phi_{\rm sum}\rangle$ can be calculated by using the distribution \eqref{steady2}. A perturbative calculation in first order in $|\epsilon|$ is enough as we are assuming weak external forces. In that case, the average field quadrature $\langle\hat{P}^\phi_{\rm sum}\rangle$ can be expressed in terms of the correlation function of the XY model with no external field ($\nu=0$) (see App. \ref{App:AppendixD}), namely
\begin{equation} \label{scaling}
\langle\hat{P}^\phi_{\rm sum}\rangle\approx\frac{2n_0|\epsilon|}{C_{\rm p}\kappa} N\sum^N_j\langle \cos(\theta_i-\theta_j)\rangle_{\epsilon=0}.
\end{equation}
The factor $N$ in Eq.\eqref{scaling} is a trivial contribution from the fact that $\hat{P}^\phi_{\rm sum}$ is a sum of $N$ copies. The correlation function 
\begin{equation}
G(|i-j|)=\langle \cos(\theta_i-\theta_j)\rangle ,
\end{equation}
on the contrary, represents a potentially non-trivial enhancement that arises from the correlations between sites. \

The importance of the correlation function in the realm of parameter estimation lies in the possible \textit{long-range order}, which in the thermodynamic limit (here $N\to\infty$) is defined as non-negligible correlations between infinitely distant sites, i.e. $\langle \cos(\theta_i-\theta_j)\rangle\neq0$ for $|i-j|\to\infty$. If this relation held, it would imply a scaling of $\sum_jG(|i-j|)$ with the system size $N$, which in turn could result in a quadratic scaling $N^2$ of $\langle\hat{P}^\phi_{\rm sum}\rangle$. This is eventually the mechanism behind the \textit{spontaneous symmetry breaking} with order parameter given by $\langle\hat{P}^\phi_{\rm sum}\rangle$, mathematically expressed as 
\begin{equation}
\lim_{\epsilon\to 0}\lim_{N\to\infty} \langle\hat{P}^\phi_{\rm sum}\rangle/N\neq 0.
\end{equation}
Nevertheless, the Mermin-Wagner theorem rules out such phase transition for a lattice dimension $D$ such that $D\leq2$ \cite{mermin66,coleman73}, where thermal fluctuations prevent ordering even at zero temperature. Particularly for a $1D$ chain, the correlation function always adopts a generic exponential decay  
\begin{equation}
G(|i-j|)\propto\exp{(-|i-j|/\xi)},
\end{equation}
where $\xi$ is the so-called \textit{correlation length} \cite{binney92}.
Even so, we propose that finite-size long-range correlations can yet be implemented in a finite chain of size $N$. This can be done by properly tuning the parameters of the lattice qubit lasers such that the correlation length becomes greater than the system size, \textit{i.e.}, $N/\xi\ll 1$, so that the correlation function gives us then an extra $N$ factor,  $\sum_j G(|i-j|)\sim N$. 
%
To this purpose, the naturally antiferromagnetic sign obtained in \eqref{steady2} does not favour this positive correlation as the ferromagnetic case does. There are two alternatives to implement an effective ferromagnetic interaction in our model; first by alternating the coupling signs $\pm t$ with the intermediate adiabatically eliminated mode so that the effective dissipative coupling becomes 
\begin{equation} \label{antiferroSign}
\mathcal{L}_{\{a_j+a_{j+1} , \frac{t^2}{\tilde{\kappa}}\}}\rightarrow \mathcal{L}_{\{a_j-a_{j+1} , \frac{t^2}{\tilde{\kappa}}\}}.
\end{equation}
Second, by alternating the phase of periodic drivings such that $\phi_j=\phi+\pi j$ to achieve the same effect expressed in \eqref{antiferroSign}.
An explicit calculation of the correlation length can be derived as the correlation functions of the classical $1D$-XY chain are well-known \cite{mattis84}.
Hence, one obtains a condition for finite-size long-range correlations in the chain,
\begin{equation}\label{finite}
	N\ln\left(\frac{I_0(4\varsigma n_0)}{I_1(4\varsigma n_0)}\right)\ll 1
\end{equation}
where $I_n(z)$ are the modified Bessel functions of the first kind. Crucially,  Eq.\eqref{finite} can be satisfied even for a weak coupling $t$ by increasing the steady number of bosons $n_0$. Notice that an increment of $n_0$ can be achieved simply by increasing the incoherent qubit pumping $\gamma$ (for a single qubit laser $n_0\approx\gamma/\kappa$). Upon condition \eqref{finite}, the quantum Fisher information for $|\epsilon|$ becomes (see App. \ref{App:AppendixE}),
\begin{equation}\label{fisher}
	F_Q[\rho_{|\epsilon|}]=\frac{2n_0N^2}{C_{\rm p}\kappa},
\end{equation}
which indicates an enhancement of $N^2$ with respect to the single qubit laser \cite{Lorenzo17}. \

An analogous procedure may be employed for estimating the phase $\phi$ for a given amplitude. Note that in this case the optimal observable for the single qubit laser, $\hat{X}_\phi$ depends itself on the target parameter $\phi$. A first estimation $\bar{\phi}$, such that $\delta\phi=(\bar{\phi}-\phi)\ll 1$, is thus required to work in the optimal operating regime. This condition is analogous to the optimal free precession time in Ramsey spectroscopy \cite{Ludlow15}. In this case the linear combination $\hat{X}^{\bar{\phi}}_{\rm sum}=\sum_j\hat{X}_j^{\bar{\phi}}$ becomes the optimal observable for the lattice qubit laser. Upon condition \eqref{finite}, the QFI becomes,
  \begin{equation}\label{fisher2}
	F_Q[\rho_{\phi}]=\frac{2n_0N^2|\epsilon|^2}{C_{\rm p}\kappa},
\end{equation}
showing again an enhancement of $N^2$. \

The results (\ref{fisher},\ref{fisher2})  both show a Heisenberg scaling $N^2$ with the number of sites, not limited by the dissipation $\kappa$. This sort of scaling is typically associated with entanglement or nonclassicality in quantum metrology \cite{giovannetti11natphys}. In contrast, here it arises solely as a result of the long-range correlations enabled by our many-body system.  It is important to notice that here the resources scale with $N$ even though we make use of many-body interactions. Yet we may achieve a quantum Fisher information scaling as $N^2$ thanks to the long-range correlations developed by the system dynamics, rather than the long-range correlations induced by a long-range interaction. This resource count is compatible with a sense of resource counting and a definition of Heisenberg limit in nonlinear estimation schemes based on the Margolus-Levitin bound \cite{zwierz10,zwierz12}. Let us recall that the $P$ function exhibits nonclassical behavior when it takes negative values or becomes more singular than the delta function \cite{Mandel95}. Here notice that the distribution \eqref{steady2} is a regular and positive function. This result thus indicates that it is possible to attain a Heisenberg scaling with classically correlated systems exhibiting long-range correlations. The natural robustness of a classical steady state renders an advantageous implementation over schemes relying on quantum states highly sensitive to decoherence. Finally, as the steady state is similar to a Gaussian state, we can safely presume that the regime in which the Cramer-Rao bound becomes valid is rapidly reached.

Let us also discriminate the roles of the key aspects involved in the results (\ref{fisher},\ref{fisher2}). In our scheme the non-unitary evolution is responsible for reaching a steady state but it is not enough to induce long-range correlations. The latter actually arise from the interplay between nearest-neighbor couplings and local many-body interactions, which
are also known to lead to a Heisenberg scaling in closed systems \cite{skotiniotis15}. 

%
%
\section{2D \& 3D systems}
Our setup benefits from having a higher dimensional lattice. 
In 2D lattices, the XY model is well-known to develop quasi-long range order for low temperatures through the Kosterlitz-Thouless transition \cite{berezinskii71,kosterlitz73,zittartz76} (2016 Nobel prize \footnote{\url{https://www.nobelprize.org/nobel_prizes/physics/laureates/2016/advanced-physicsprize2016.pdf}}). This transition is driven by the energy cost to thermally break up pairs of vortex-antivortex configurations. The critical temperature is approximately located at $\beta_{\rm c}\varsigma=2/\pi$. The effective temperature $\beta_{\rm eff}=n_0/(C_{\rm p}\kappa)$ implies that low temperatures are achieved close the the critical point ($C_{\rm p}\approx 1$) and large average number of bosons $n_0$. Hence, our regime of parameters readily guarantees that we work in an effective low temperature regime $n_0\varsigma>2/\pi$. In this regime, the correlation function in Eq.\eqref{scaling} decays algebraically, \textit{i.e,} 
\begin{equation}
G(|i-j|)\propto|i-j|^{-\eta},
\end{equation}
which softens the condition imposed for achieving finite-size long-range correlations. Specifically, by using the spin wave approximation for the value of $\eta$ \cite{zittartz76}, we have
\begin{equation}\label{finite2}
N^{\frac{1}{2\pi n_0\varsigma}}\approx 1.
\end{equation}

In 3D lattices the XY model undergoes Spontaneous Symmetry Breaking and it naturally shows long-range order. Consequently conditions (\ref{finite},\ref{finite2}) are not necessary to achieve the enhancing $N^2$. 

%

\section{Conclusions}
The model introduced in \eqref{Liouvillian} may be implemented with single qubit photon laser using single atoms \cite{mckeever03} or superconducting qubits \cite{astafiev07nat,Hauss08,navarrete14prl}. The phononic excitations in ion traps can also play the role of the bosonic field \cite{vahala09,grudinin10}, in which case this systems allows the precise measurement of ultra-weak forces resonant with the trapping frequency \cite{Lorenzo17,biercuk10natnano,schreppler14sci,Ivanov13pra,shaniv16}. A possible implementation of our model with trapped ions may be carried out by extending the implementation sketched in \cite{Lorenzo17}, in which it was shown that local sources of error such as heating or dephasing only result in a renormalization of the parameters.  \

The ideas exposed do not fundamentally rely on the XY model. They can be readily generalized to other setups that give rise to a dissipative dynamics in which the steady state can be formally identified in terms of another classical Hamiltonian $H_{\rm c}$ of the form 
\begin{equation}
H_{\rm c}=\varphi\sum_j^NH_j+t\sum_{\langle j,k\rangle}H_{i,j},
\end{equation}
like Eq.\eqref{steady1}. Here we are only assuming short-range interactions, implied by the notation $\langle j,k\rangle$, so that the resources (number of probes) scale with $N$. If the target parameter $\varphi$ is small enough, the Fisher information of the steady state will generically be expressed as 
\begin{equation}
F_Q[\rho_\varphi]\propto N\sum_j G(|i-j|),
\end{equation}
with $G$ being the corresponding correlation function of the equivalent classical model $H_{i,j}$. Bearing in mind the fluctuation-dissipation theorem \cite{binney92}, this establishes a general link between a susceptibility $\chi\propto \sum_j G(|i-j|)$ and the Fisher information in the steady state, thereby showing the possibility of a metrological enhancement though long-range correlations in dissipative systems. An analogous result was recently exposed in \cite{martinez16}, and it strengthens a connection between the fields of quantum metrology and condensed-matter Physics.

\section{Acknowledgments} Funded by the People Programme (Marie Curie
Actions) of the EU’s Seventh Framework Programme under REA Grant Agreement no: PCIG14-GA-2013-630955. We thank Pedro Nevado for fruitful discussions.

\appendix
\section{Mean field theory} \label{App:AppendixA}
Here we shall perform a mean-field analysis of equation \eqref{Liouvillian}, in the spirit of the well-known Maxwell-Bloch equations of a laser \cite{BreuerPet}. The mean-field ansatz assumes that the system density matrix $\rho$ is separable in the qubit-field subspaces, \textit{i.e.}, $\rho\approx \rho_{\rm field}\otimes\rho_{\rm qubit}$. In practical terms, this allows us to approximate expectation values in such a way that $\langle \sigma a \rangle\approx \langle\sigma\rangle\langle a\rangle$. Furthermore, this avoids the infinite hierarchy of equations for the expectation values of moments of such observables. Namely, we can write the following closed system of equations in terms of the variables $A_j\equiv\langle a_j\rangle$, $S_j\equiv-i\langle \sigma_j^- \rangle$ and $D_j\equiv\langle \sigma^z_j \rangle$,

\begin{align}
\dot{A}_j &=-gS_j-C A_j -\frac{t^2}{\tilde{k}}(A_{j-1}+A_{j+1}) \label{meanfieldEqs}\\ 
\dot{S}_j &=gD_jA_j -\gamma S_j \nonumber\\ 
\dot{D}_j &=-2g(S^*_jA_j+S_jA^*_j)-2\gamma(D_j-1) \nonumber ,
\end{align}
where periodic boundary conditions are assumed. The result \eqref{meanfieldEqs} is achieved by means of the Heisenberg equations for such observables and using the commutation relations $[a^{\dag},a]=1$, $\left\{\sigma^+,\sigma^-\right\}=1$ and $[\sigma^z,\sigma^{\pm}]=\pm 2 \sigma^{\pm}$. The set of nonlinear equations \eqref{meanfieldEqs} represents an extension of the Maxwell-Bloch with an extra term describing the hoping of bosons between sites.	This system of equations exhibits multiple and frequently complicated possible steady states depending on the regime of parameters studied. It is noticeable also that one can find chaotic behavior in some regions of the parametric space given by $\kappa,g,t,\gamma$. This should not be surprising as the Maxwell-Block equations can be shown to be equivalent to the well-known Lorentz equations \cite{Mandel95}. Therefore, appropriate ansatzs for the steady state must be assumed for a certain regime of parameters. \

We shall consider in the following that the laser operates in a regime such that the pumping of the qubits represents the smallest timescale in the problem, \textit{i.e.}, $\gamma >> \kappa, g, t$, which is consistent with the adiabatic elimination of the qubits. In this case, the fast variables $S_j$ and $D_j$ may be adiabatically eliminated to obtain an equation for $A_j$. Additionally, we shall assume that the system does not break the translation symmetry, hence $A_j=A$. In writing Eq. \eqref{meanfieldEqs} in a basis of the chain normal modes with $A_{q}=\sum_j A_j e^{iqj}$, the only surviving mode is the fundamental mode $q_0=0$, hence $A_{q_0}=\sum_j A_j$. After using the adiabatic elimination, the equation for this mode adopts the form,
\begin{equation} \label{mfmode}
\frac{dA}{dt}=\left(\frac{NC_{\rm p}}{1+\frac{|A|^2}{n_{\rm mf}}}-(C+\tilde{t})\right)A,
\end{equation}
with $C_{\rm p}=g^2/\kappa\gamma$ and $n_{\rm mf}=2\gamma^2/g^2$. Equation \eqref{mfmode} exhibits a Hopf bifurcation indicating a dissipative phase transition to a lasing phase for 
\begin{equation}
\tilde{C}_{\rm p}=\frac{C_{\rm p}}{\left(1+3\left(\frac{t}{k}\right)^2\right)}>1,
\end{equation}
which has the stable solution $|A_{q_0}|^2=Nn_{\rm mf}(\tilde{C}_p-1)$. This result simply represents a renormalization of the pumping parameter $\tilde{C_{\rm p}}$ with respect to the single qubit laser ($t=0$), in which the critical point is given by $C_{\rm p}=1$ \cite{BreuerPet,Lorenzo17}.
\section{Adiabatic elimination}\label{App:AppendixB}
In this section we shall derive the effective master equation claimed in equation \eqref{adiabatic}. We shall use a straightforward generalization of the procedure employed for the single-qubit laser \cite{Lorenzo17}. Firstly, we trace over the qubits from the master equation \eqref{Liouvillian},
\begin{multline}
\dot{\rho}_f=-i g\sum^N_j(a_j \rho^{ge}_j +a_j^{\dag} \rho^{eg}_j -\rho^{ge}_j a_j -\rho^{eg}_ja^{\dag}_j) - \\
	-i\sum_j^N (\epsilon a_j^{\dag} \rho_f +\epsilon^* a_j \rho_f -\epsilon\rho_fa_j^{\dag} -\epsilon^*\rho_f a_j^{\dag})+ \\
	+(\kappa +\frac{t^2}{\tilde{\kappa}})\sum_j^N (2a_j \rho_f a_j^{\dag} - a^{\dag}_j a_j \rho_f -\rho_fa^{\dag}_j a_j) + 	\\
	+\frac{t^2}{\tilde{\kappa}}\sum_j (2a_j \rho_fa_{j+1}^{\dag}  - a_{j+1}^{\dag}a_j \rho_f -\rho_fa_{j+1}^{\dag}a_j +c.c) , \label{rhof}
\end{multline}
where we introduced the notation $\rho^{ge}_j=\langle g_j| \rho|e_j \rangle$ and $\epsilon=|\epsilon|e^{i\phi}$. In order to obtain a closed equation for the reduced density matrix $\rho_f$, we have to eliminate the operators $\rho^{ge}_j,\rho^{eg}_j$ from Eq. \eqref{rhof}. By writing their corresponding equations of motion using Eq. \eqref{Liouvillian},
\begin{equation}
\dot{\rho}^{ge}_j=-i g(a_j^{\dag}\rho^{ee}_j -\rho^{gg}_ja_j^{\dag}) - \gamma \rho^{ge}_j ,\label{rhoge}
\end{equation}
(where we have neglected the contributions from $\kappa$,$\epsilon$ and $t^2/\tilde{\kappa}$ in comparison with $\gamma$) the operators $\rho_{ge}$ and $\rho_{eg}$ can be adiabatically eliminated (in the limit  $\gamma \gg \kappa, g, |\epsilon|$) from \eqref{rhof} by taking $\dot{\rho}_{ge}\approx 0$ in Eq. \eqref{rhoge}  and substituting in Eq. \eqref{rhof} their steady-state solutions,
\begin{equation}
\rho^{ge}_j=-i \frac{g}{\gamma}(a_j^{\dag} \rho^{ee}_j-\rho^{gg}_j a_j^{\dag}) . \label{rhoeg}
\end{equation}
Likewise, the equations of motion of the operators $\rho_{gg}$ and $\rho_{ee}$ are required as the resulting equation still depends on them. These can be derived again from Eq. \eqref{Liouvillian}, namely
\begin{align}
\dot{\rho}^{ee}_j &=-i g(a_j \rho^{ge}_j-\rho^{eg}_j a_j^{\dag}) +2\gamma \rho^{gg}_j \label{rhoeedot}\\ 
\dot{\rho}^{gg}_j &=-i g(a_j^{\dag} \rho^{eg}_j-\rho^{ge}_j a_j) -2\gamma \rho^{gg}_j \label{rhoggdot}
\end{align}
where we again neglect terms with $\kappa$, $\epsilon$ and $t^2/\tilde{\kappa}$. A perturbative solution to the steady-states of Eqs. (\ref{rhoeedot},\ref{rhoggdot}) may now be expressed in terms of the field density matrix $\rho_f$. To do so, let us adiabatically eliminate $\rho^{gg}_j$ by taking $\dot{\rho}^{gg}_j\approx 0$ in Eq.\eqref{rhoggdot}, which gives
\begin{multline}
\rho^{gg}_j= -\frac{ig}{2\gamma}(a^{\dag}_j \rho^{eg}_j-\rho^{ge}_j a_j)=  \label{rhogg} \\
=\frac{g^2}{2\gamma^2}(2a_j^{\dag} \rho^{ee}_ja_j-a{\dag} a_j \rho^{gg}_j-\rho^{gg}_ja_j^{\dag} a_j).
\end{multline}
The ground state population of each qubit is expected to be negligible as a result of the fast pumping of the qubits ($\gamma \gg 1$). Thus, in first order we can assume $\rho^{gg}_j\approx 0$ and $\rho^{ee}_j=\rho_j-\rho^{gg}_j\approx \rho_j$. A second order correction is achieved by inserting this first order approximation into Eq.\eqref{rhogg}, hence
\begin{align}
\rho^{gg}_j&= \frac{g^2}{\gamma^2}\adag \rho_j a \label{rhogg2}\\ 
\rho^{ee}_j&=\rho_j-\rho^{gg}_j=\rho_j- \frac{g^2}{\gamma^2}a^{\dag} \rho_j a_j \label{rhoee}.
\end{align}
A closed equation for $\rho_f$ is finally accomplished by inserting Eqs.(\ref{rhogg2},\ref{rhoee}) into Eq.\eqref{rhof} and bearing in mind that $\sum_j a_j\rho_j=\sum_j a_j\rho_f$
\begin{multline}
	\dot{\rho}_f= -i \sum_j(\epsilon a_j^{\dag} \rho_f +\epsilon^* aj \rho_f -\epsilon\rho_fa_j^{\dag} -\epsilon^*\rho_f a_j)+ \\ 
		+\frac{g^2}{\gamma}\sum_j( 2 a_j^{\dag} \rho_f a_j -a_ja_j^{\dag} \rho_f- \rho_f a_ja_j^{\dag}) + \\
		+\frac{2g^4}{\gamma^3}\sum_j(  a_ja_j^{\dag} \rho_f a_j a_j^{\dag}-{a_j^{\dag}}^2 \rho_f a_j^2) + \\
		+(\kappa +\frac{t^2}{\tilde{\kappa}})\sum_j^N (2a_j \rho_f a_j^{\dag} - a^{\dag}_j a_j \rho_f -\rho_fa^{\dag}_j a_j) + 	\\
	+\frac{t^2}{\tilde{\kappa}}\sum_j (2a_j \rho_fa_{j+1}^{\dag}  - a_{j+1}^{\dag}a_j \rho_f -\rho_fa_{j+1}^{\dag}a_j +c.c) . \label{adiabaticEquation}
\end{multline}
which, written in compact notation, is the result presented in equation \eqref{adiabatic}.\

\section{Fokker-Planck equation}\label{App:AppendixC}
Let us summarize the derivation of the Fokker-Planck equation \eqref{PFokker1}, the angular Fokker-Planck equation \eqref{PFokker2} and their corresponding steady-state solutions (\ref{steady1},\ref{steady2}). Bearing in mind the coherent representation of a density matrix $\rho$,
\begin{equation}\label{coherentP}
\rho(t)=\int d^2\alpha P(\alpha,\alpha^*,t) |\alpha\rangle\langle \alpha |, 
\end{equation}
the master equation can be expressed as an equation of motion for $P(\alpha,\alpha^*,t)$ after an integration by parts with the assumption of zero boundary conditions at infinity. Note that this change introduces an extra minus sign for each differential operator $\partial_{\alpha}$. The integrand of Eq.\eqref{coherentP} is hence expressed as a product of $ |\alpha\rangle\langle \alpha |$ and a \textit{c}-number function of $\alpha,\alpha^*$, yielding a differential equation for $P(\alpha,\alpha^*,t)$. We shall focus in a regime in which the average number of bosons is large, which implies $|\alpha|^2\gg 1$. As $B$ is a very small coefficient compared to $A$, $B/A\propto(g/\gamma)^2 \ll 1$, we shall retain only the most important terms in $B$ and drop any contribution smaller than $B|\alpha|^2 \alpha$. By doing so, we arrive at the Fokker-Planck equation claimed in equation \eqref{PFokker1},
\begin{align} \label{PFokker1A}
\frac{\partial P}{\partial t}&=+2A\sum_j\frac{\partial^2P}{\partial \alpha_j\partial\alpha_j^*} \\
 &-\sum_{\langle j,k\rangle}\frac{\partial}{\partial \alpha_j} [(A-C-B|\alpha_j|^2)\alpha_j  -D\alpha_k-\epsilon'] P +c.c. , \nonumber
\end{align}
where $\epsilon'\equiv i\epsilon$. Let us rewrite equation \eqref{PFokker1A} in cartesian coordinates, with $\alpha_j=x^1_j+i x^2_j$ and $\partial/\partial\alpha=1/2(\partial/\partial x^1- i\partial/\partial x^2)$,
\begin{align} \label{PFokker2A} 
\frac{\partial P}{\partial t}&= \frac{A}{2}\sum^N_j\sum^2_{i=1}\frac{\partial^2P}{\partial x^2_i} \\
&-\sum^N_j\sum^2_{i=1}\frac{\partial}{\partial x^i_j} [(A-C-B \vec{x_j}^2)x^i_j  -D(x^i_{j-1}+x^i_{j+1})-\epsilon'_i] P  \nonumber
\end{align}
where we introduced the two-dimensional vectors $\vec{x}=(x^1_j,x^2_j)$ and $\vec{\epsilon'}=(\Re(\epsilon'),\Im(\epsilon'))$. The steady-state satisfies $\partial P/\partial t=0$ and Eq.\eqref{PFokker2A} can be written as $\sum_i\partial J_i/\partial x_i=0$, with the current $\vec{J}_j$ defined as
\begin{equation}
J^i_j=  [(A-C-B \vec{x_j}^2)x^i_j  -D(x^i_{j-1}+x^i_{j+1})-\epsilon'_i]-\frac{A}{2}\frac{\partial P}{\partial x^i_j}.
\end{equation}
Fortunately, the drift vector $A^i_j\equiv [(A-C-B \vec{x}^2)x^i_j -D(x^i_{j-1}+x^i_{j+1}) -\epsilon'_i]$ satisfies a detailed balance condition given by $\partial A_i/\partial x_j=\partial A_j/\partial x_i$, and the steady-state solution can be hence found by the condition $\vec{J}=0$ \cite{Mandel95}. This gives rise to a first order differential equation for $P$ that can be trivially integrated to give,
\begin{equation}\label{PestA}
\begin{split} 
P(\vec{x})= \frac{1}{Z} \exp  &  \left\{\frac{1}{A}\left[\left(A-C-\frac{B}{2}\vec{x_j}^2\right)\vec{x_j}^2\right]- \right. \\
  &\quad \left. {} \frac{2}{A}\left[D\sum_{\langle j,k\rangle}\vec{x_j}\cdot\vec{x_k}-\vec{\epsilon'}\cdot\vec{x_j}\right]    \vphantom{}\right\},
\end{split}
\end{equation}
where $Z$ is a normalization constant. This can be then expressed in polar coordinates $\alpha_j=r_je^{i\theta_j}$ as follows,
\begin{equation}\label{Pest2A}
\begin{split} 
P(\vec{r},\vec{\theta})= \frac{1}{Z} \exp  & \left( \sum_j(\mu r_j^2-\lambda r_j^4  -2\nu r_j\sin{(\theta_j-\phi)} )- \right. \\
  &\quad \left. {} -\sum_{\langle j,k\rangle}2\varsigma r_jr_k \cos{(\theta_j-\theta_k)}  \vphantom{}\right) ,
\end{split}
\end{equation}
where we introduced the notation  $\vec{r}=(r_1,r_2,\cdots,r_N)$ and $ \vec{\theta}=(\theta_1,\theta_2,\cdots,\theta_N)$, and defined the parameters 
$\lambda = B/2A$,
$\mu=(A-C)/A$,
$\nu=|\epsilon|/A$
 and $\varsigma=D/A$.\

One can derive from an equation solely for the angular variables $\vec{\theta}$ from Eq. \eqref{PFokker1A}. To do so, on has to admit the radial variables are settled around their steady-state values $r_j\approx r_0$, while the dynamics of Eq. \eqref{PFokker1A} is hence dominated by the angular components. In that case, the $P$ function may be assumed to take the form $P(\mathbf{r},\mathbf{\theta})=R(r_1)R(r_2)\cdots R(r_N)P'(\mathbf{\theta})$ where each $R(r_j)$ is a properly normalized Gaussian function around $r_0$, 
\begin{equation}\label{radial}
R(r_j)=\frac{1}{N}\exp{\left(-\frac{(r_j-r_0)^2}{2\sigma^2}\right)} 
\end{equation}
Above threshold in a regime of large number of bosons, the normalization constant in \eqref{radial} is given by
\begin{align}\label{normaR}
N&=\int_0^{\infty}rdr\exp{\left(-\frac{(r_j-r_0)^2}{2\sigma^2}\right)} =\\
&r_0\int_{-r_0\approx-\infty}^{\infty}dr'\exp\left(-\frac{r'^2}{2\sigma^2}\right)+\cancel{\int_{-r_0}^{\infty}rdr\exp{\left(-\frac{r'^2}{2\sigma^2}\right)}} = \\
&=r_0\sqrt{2\pi\sigma^2}.
\end{align}
Our equations can be written in polar coordinates with the aid of the equivalences,
\begin{align}
\frac{\partial}{\partial\alpha}&=\frac{1}{2}e^{(-i\theta)}\left(\frac{\partial}{\partial r}-\frac{i}{r}\frac{\partial}{\partial}\right) \nonumber\\
\frac{\partial}{\partial\alpha*}&=\frac{1}{2}e^{(i\theta)}\left(\frac{\partial}{\partial r}+\frac{i}{r}\frac{\partial}{\partial}\right).
\end{align}
The equation \eqref{PFokker1A} then reads as follows,
\begin{align}\label{FokkerPolarA}
&\frac{\partial P}{\partial t}= \frac{\prod_j\partial R_j}{\partial t}P'+\frac{\partial P'}{\partial t}\prod_j R_j= +\frac{A}{2}\sum_j\frac{\partial^2}{\partial \theta_j^2}P- \nonumber\\
&-\sum_j\left\{\frac{1}{r_j}\frac{\partial}{\partial r_j}\left[r_j^2(A-C-Br_j^2)P\right] +\frac{A}{2}\left[\frac{\partial^2}{\partial r_j^2}+\frac{1}{r_j^2}\frac{\partial}{\partial r_j}\right]P\right\}+\nonumber\\
&+|\epsilon|\sum_j\sin(\theta_j-\phi)\frac{\partial P}{\partial r_j}+\sum_j\frac{|\epsilon|}{r_j}\cos(\theta_j-\phi)\frac{\partial P}{\partial\theta_j}+\nonumber\\
&+D\sum_{\langle j,k\rangle}r_k\cos(\theta_k-\theta_j)\frac{\partial P}{\partial r_j}+D\sum_{\langle j,k\rangle}\frac{r_k}{r_j}\sin(\theta_k-\theta_j)\frac{\partial P}{\partial\theta_j}.
\end{align}
One can obtain a purely angular equation by integrating both sides of Eq. \eqref{FokkerPolarA} in the radial variables $\int_0^{\infty}\vec{r}d\vec{r}$. On the one hand, the second line in \eqref{FokkerPolarA} can be simplified (for $|\epsilon|\approx 0$) as $R(r_j)$ satisfies
\begin{align}
&\frac{\partial R_j}{\partial t}= \\
&-\left\{\frac{1}{r_j}\frac{\partial}{\partial r_j}\left[r_j^2(A-C-Br_j^2)R_j\right] +\frac{A}{2}\left[\frac{\partial^2}{\partial r_j^2}+\frac{1}{r_j^2}\frac{\partial}{\partial r_j}\right]R_j\right\} .\nonumber
\end{align}
On the other hand, the integration of the first derivative $\partial_{r}R$ is eliminated through the relation,
\begin{align}
\int_0^{\infty}rdr\partial_rR&=-\frac{1}{N}\int_0^{\infty}rdr\frac{(r-r_0)}{\sigma^2}\exp\left(-\frac{(r-r_0)^2}{2\sigma^2}\right)=\nonumber\\
&=-\frac{1}{N}\int_{-r_0}^{\infty}(r'+r_0)dr'\frac{(r')}{\sigma^2}\exp\left(-\frac{(r')^2}{2\sigma^2}\right)\approx\nonumber\\
&\approx-\frac{1}{N}\int_{-\infty}^{\infty}dr'\frac{(r'^2)}{\sigma^2}\exp\left(-\frac{(r')^2}{2\sigma^2}\right)=\nonumber\\
&=-\frac{1}{N}\sqrt{2\pi}\sigma=-\frac{1}{r_0}.
\end{align}

After grouping terms, the resulting equation adopts the form claimed in equation \eqref{PFokker2},
\begin{align} \label{PFokker3A}
\frac{\partial P'}{\partial t}&=+\frac{A}{2n_0}\sum_j\frac{\partial^2P'}{\partial \theta_j^2}  \\
&+\sum_{\langle j,k\rangle}\frac{\partial}{\partial \theta_j}\left((D\sin(\theta_k-\theta_j)+\frac{|\epsilon|}{\sqrt{n_0}}\cos(\theta_j-\phi))P'\right) \nonumber,
\end{align}
The steady state state of \eqref{PFokker3A} can be obtained by imposing a detail balance condition such that the current $\vec{J}_{\theta}=0$ or simply by taking $r_j=r_0$ in \eqref{PestA} and grouping the radial part into the normalization constant $Z$, which gives
\begin{equation}\label{steady2A}
\begin{split} 
P'(\vec{\theta})= \frac{1}{Z} \exp   \left(  -\sum_{\langle j,k\rangle}2\varsigma  n_0 \cos{(\theta_j-\theta_k)}- \right. \\
  \quad \left. {} -\sum_j 2\nu \sqrt{n_0}\sin{(\theta_j-\phi)}  \vphantom{}\right) .
\end{split}
\end{equation}
%
\section{Correlation function in the XY chain} \label{App:AppendixD}
In this section we aim to show rigorously the expression of $\langle\hat{P}^\phi_{\rm sum}\rangle=\langle\sum_j\hat{P}^\phi_j\rangle$ in first order in $\epsilon$ as claimed in equation \eqref{scaling} as well as the expresion for $\langle\hat{X}^{\bar{\phi}}_{\rm sum}\rangle=\langle\sum_j\hat{X}^\phi_j\rangle$. Concretely, we will show how it can be written in terms of the correlation function of the classical XY model with no external field.
The correlation functions of the XY chain are already well-known \cite{mattis84}. In particular, the only two-point non-zero correlation function of the Boltzmann distribution \eqref{Pest2A} with $\nu=0$ is precisely $\langle \cos(\theta_i-\theta_j)\rangle$ which can expressed in terms of the modified Bessel functions of the first kind $I_n(z)$, namely
\begin{equation}\label{correlationF}
\langle \cos(\theta_1-\theta_{j+1}\rangle=\frac{1}{Z_0}\sum_{n=-\infty}^{\infty}I_{n-1}^{j}(4r_0^2\varsigma)I_{n}^{N-j}(4r_0^2\varsigma).
\end{equation}
In Eq.\eqref{correlationF} we have assumed ferromagnetic sign. For distant sites, the correlation function in 1D systems is known to decay exponentially with a certain correlation length $\xi$ (the typical scale of the correlations) \cite{binney92}, i.e.,
\begin{equation}
\langle \cos(\theta_1-\theta_{1+j}\rangle\approx\left(\frac{I_1(4r_0^2\varsigma)}{I_0(4r_0^2\varsigma)}\right)^j\approx e^{-j/\xi},
\end{equation}
from which the correlation length is given by
\begin{equation}
\xi^{-1}=\ln\left(\frac{I_1(4r_0^2\varsigma)}{I_0(4r_0^2\varsigma)}\right).
\end{equation}

A perturbative expression in first order of $\epsilon$ for $\hat{P}^\phi_{\rm sum}=\sum_j\hat{P}^\phi_j$ can be derived by using the Boltzmann factor given by the angular $P$ function calculated in Eq.\eqref{Pest2A}. If we expand the exponential up to first order in $\nu$, the average quadrature $\langle\hat{P}^\phi_{\rm sum}\rangle$ is given by two contributions $\langle\hat{P}^\phi_{\rm sum}\rangle\approx{\langle\hat{P}^\phi_{\rm sum}\rangle}_0+\delta\langle\hat{P}^\phi_{\rm sum}\rangle$ in terms of the Boltzmann factor of the XY with no external field,i.e.,
\begin{equation}
P_0(\vec{\theta})= \frac{1}{Z_1} \exp\left(4\varsigma\sum_j r_jr_{j+1}\cos(\theta_j-\theta_{j+1}) \right) 
\end{equation}
with $Z1$ being the partition function up to first order, $Z_1=Z_0+\delta Z$. It is straightforward to check that the first order contribution $\delta Z$ is zero so $Z_0=Z_1$. The zero order contribution is then,
\begin{equation}
{\langle\hat{P}^\phi_{\rm sum}\rangle}_0=\frac{1}{Z_0}\oint d\vec{\theta}(-2\sum_jr_0\sin(\theta_j-\phi))P_0(\vec{\theta})
\end{equation}
while the first order contribution can be expressed as,
\begin{align}
&\delta{\langle\hat{P}^\phi_{\rm sum}\rangle}= \\
&\frac{1}{Z_0}\oint d\vec{\theta}(-2\sum_jr_0\sin(\theta_j-\phi))(-2\nu\sum_kr_0\sin(\theta_k-\phi))P_0(\vec{\theta}). \nonumber
\end{align}
The average quadrature is thus given by
\begin{equation} \label{PquadratureA}
\langle\hat{P}^\phi_{\rm sum}\rangle\approx 4 r_0^2\nu\sum_{i,j}\langle \sin(\theta_j-\phi)\sin(\theta_i-\phi)\rangle_{\epsilon=0}.
\end{equation}
By using the trigonometric relation,
\begin{equation}\label{trigon1}
\sin(\theta_j-\phi)\sin(\theta_i-\theta_j)=\frac{1}{2}(\cos(\theta_i-\theta_j)-\cos(\theta_i+\theta_j+2\phi)),
\end{equation}
we note that only the first term in Eq.\eqref{trigon1} gives rise to a non-zero correlation contribution, thus the quadrature takes the form claimed in equation \eqref{scaling},
\begin{equation} \label{scalingA}
\langle\hat{P}^\phi_{\rm sum}\rangle\approx\frac{2n_0|\epsilon|}{C_{\rm p}\kappa} N\sum^N_j\langle \cos(\theta_i-\theta_j)\rangle_{\epsilon=0}.
\end{equation}

The field quadrature $\hat{X}^{\bar{\phi}}_{\rm sum}=\sum_j\hat{X}^\phi_j$ in first order, on the other hand, will be given by
\begin{equation}\label{Xquadrature}
\langle\hat{X}^{\bar{\phi}}_{\rm sum}\rangle\approx 4 r_0^2\nu\sum_{i,j}\langle \cos(\theta_j-\bar{\phi})\sin(\theta_i-\phi)\rangle_{\epsilon=0}.
\end{equation}
As we assume $\delta\phi=(\bar{\phi}-\phi)\ll 1$, Eq.\eqref{Xquadrature} can be further simplified by means of the trigonometric relation
\begin{align}\label{trigon2}
\sin(\theta-\bar{\phi})&=\cos(\theta-\phi)\cos{\delta\phi}+\sin(\theta-\phi)\sin\delta\phi\approx \nonumber\\
&\approx\cos(\theta-\phi)+\sin(\theta-\phi)\delta\phi.
\end{align}
Only the second term in \eqref{trigon2} leads to a non-zero correlation function, so we finally arrive to
\begin{align}
\langle\hat{X}^{\bar{\phi}}_{\rm sum}\rangle&\approx \delta\phi4 r_0^2\nu\sum_{i,j}\langle \sin(\theta_i-\phi)\sin(\theta_j-\phi)\rangle_{\epsilon=0}=\nonumber\\
&= \delta\phi4 r_0^2\nu\sum_{i,j}\langle \cos(\theta_i-\theta_j)\rangle_{\epsilon=0}.\label{Xquadrature2}
\end{align}
By imposing the condition $N/\xi\ll 1$, i.e.
\begin{equation}\label{finiteA}
	N\ln\left(\frac{I_0(4\varsigma n_0)}{I_1(4\varsigma n_0)}\right)\ll 1 ,
\end{equation}
relations (\ref{PquadratureA},\ref{Xquadrature2}) are further reduced,yielding
\begin{align}
\langle\hat{P}^\phi_{\rm sum}\rangle&\approx 4 n_0\nu N^2\label{Pfinal}\\
\langle\hat{X}^{\bar{\phi}}_{\rm sum}\rangle&\approx \delta\phi4 n_0\nu N^2,\label{Xfinal}
\end{align}
where $n_0=r_0^2$ symbolizes the steady average of bosons of each cavity.
\section{Symmetric logarithmic derivative \& quantum Fisher information} \label{App:AppendixE}
In this section we will obtain the optimal observables for the lattice-qubit laser, as well as the quantum Fisher information for them. To do that, we have to solve the operator equation for the symmetric logarithmic derivative, this is,
\begin{equation}
 \partial_\varphi \rho_\varphi=\frac{1}{2}(\rho_\varphi L_\varphi+L_\varphi \rho_\varphi). \label{SLD}
\end{equation}
The $P$ function \eqref{Pest2A} can be well approximated by the following Gaussian-like approximation (we treat directly the ferromagnetic case),
\begin{equation}\label{PpolarGauss}
\begin{split} 
P(\vec{r},\vec{\theta})= \frac{1}{Z} \exp  & \left( -\sum_j\frac{(r_j-r_0)^2}{2\sigma^2}- \nu\sum_j r_j\sin{(\theta_j-\phi)}+ \right. \\
  &\quad \left. {} +4\varsigma\sum_j r_jr_{j+1}\cos(\theta_j-\theta_{j+1})  \vphantom{}\right) ,
\end{split}
\end{equation}
where the radial components are assumed to be settled around their steady-state values $r^2_0$ with width $\sigma^2$. Using \eqref{PpolarGauss}, the l.h.s of Eq.\eqref{SLD} reads,
\begin{equation}
\partial_{|\epsilon|}P(r,\theta)=  \left(\partial_{|\epsilon|}\ln Z+\frac{i}{A}\sum_j(\alpha_j e^{-i\phi}-\alpha_j^*e^{i\phi})\right)P \label{derivP}.
\end{equation}
It is straightforward to check that $\partial_{|\epsilon|}\ln Z$ is equivalent to the average of the sum of the field quadratures  $\langle \hat{P}_\phi \rangle=\langle i\sum_j(a_je^{-i\phi}-a_j^{\dag}e^{i\phi}) \rangle$. This result suggests we introduce the ansatz $L_{|\epsilon|}=S_0+\sum_j(S a_j+S^*a_j^{\dag})$, with $S_0,S$ proper coefficients. Inserting this ansatz into the r.h.s of Eq.\eqref{SLD} and using the relations \eqref{alphaderiv}\eqref{alphaderiv2}, we obtain
\begin{equation}
L_{|\epsilon|}\rho=\int_{0}^\infty\int_{0}^{2\pi} d\vec{\theta} d\vec{r}r_1\ldots r_N(S_0+\sum_j(S\alpha_j+S^*(\alpha_j^*-\partial_{\alpha_j})))P \label{ansatz}
\end{equation}
with analogous expression for $\rho L_{|\epsilon|}$. Well above threshold where $r_0 \gg \sigma$, we may further simplify the exact derivative $\partial_\alpha$ in Eq.\eqref{ansatz} if we assume the radial component to be approximately constant and homogeneous so that $\alpha_j=r_je^{i\theta_j}\approx r_0e^{i\theta_j}$ (\textit{after} taking the derivative). We may distinguish two contributions to the derivative. First, an on-site contribution given by the first two terms of the r.h.s in Eq.\eqref{PpolarGauss}
\begin{multline} \label{onsite}
\partial_{\alpha_j} P_{\rm on}=\frac{e^{-i\theta_j}}{2}(\frac{\partial}{\partial r_j}-\frac{i}{r_j}\frac{\partial}{\partial\theta_j})P_{\rm on}= \\
=(-\frac{\alpha_j^*}{2\sigma^2}+\frac{r_0}{2\sigma^2}e^{-i\theta_j}+\frac{i|\epsilon|}{A}e^{-i\phi})P\approx  \frac{i|\epsilon|}{A}e^{-i\phi}P_{\rm on} .
\end{multline}
Second, a contribution given by the neighboring interaction,
\begin{equation} \label{interaction}
\partial_{\alpha_j} P_{\rm int}\approx 2\varsigma r_0(e^{-i\theta_{j-1}}+e^{-i\theta_{j+1}})P_{\rm int}.
\end{equation}
Using the results (\ref{ansatz},\ref{onsite},\ref{interaction}), we can identify terms from both sides of the equation \eqref{SLD}, leading to the following SLD,
\begin{equation}\label{SLDepsilon}
 L_{\epsilon}[\rho_{|\epsilon|}]=\frac{1}{A}\left(\frac{N\nu}{1-2\varsigma}-\sum_j\langle \hat{P}^\phi_j\rangle + \frac{\sum_j\hat{P}^\phi_j}{1-2\varsigma} \right) 
\end{equation}
The contribution of the first term of the r.h.s can be neglected in comparison with the contribution given by $\sum_j\hat{P}_\phi$. On the other hand, the SLD must fulfill the relation $\langle L_{|\epsilon|} \rangle=0$ according to the definition \eqref{SLD}. In our case, this implies that the result \eqref{SLDepsilon} is correct as long as $\varsigma\ll 1$, which is consistent with our scheme. In that case, the observable $\hat{P}^\phi_{\rm sum}=\sum_j\hat{P}^\phi_j$ turns out to be the optimal observable for estimating $|\epsilon|$. A totally analogous procedure may be employed to prove that $\hat{X}^\phi_{\rm sum}=\sum_j\hat{X}^\phi_j$ is the optimal observable for estimation $|\phi|$.\

As the quantum Fisher information is obtained through the SLD, we may recover the results claimed in equations (\ref{fisher},\ref{fisher2}) simply through the error propagation formula, for which the fluctuations $\Delta^2\langle\hat{P}^\phi_{\rm sum}\rangle$, $\Delta^2\langle\hat{X}^{\bar{\phi}}_{\rm sum}\rangle$ are additionally needed. These fluctuations can be written as,
\begin{equation}\label{fluctuations}
\Delta^2\langle\hat{P}^\phi_{\rm sum}\rangle=\sum_{i,j}\langle\hat{P}^\phi_i\hat{P}^\phi_j\rangle-\langle \sum_j\hat{P}^\phi_j\rangle^2
\end{equation}
and analogously for $\Delta^2\langle\hat{X}^{\bar{\phi}}_{\rm sum}\rangle$. Notice that the thermal averages in \eqref{fluctuations} are \textit{not} at zero external field ($\nu\neq 0$). The second term in Eq.\eqref{fluctuations} is directly given by \eqref{Pfinal}, which can be neglected as it leads to a second order contribution in $|\epsilon|$. The first term in turn may be straightforwardly derived with the aid of the following relation held by the partition function $Z$,
\begin{equation}
\frac{1}{Z}\frac{\partial^2 Z}{\partial\phi^2}=-\nu\sum_j\langle\hat{P}^{\phi}_{j}\rangle+\nu^2\sum_{i,j}\langle\hat{X}^\phi_i\hat{X}^\phi_j\rangle.
\end{equation}
The partition function $Z$ is not expected to explicitly depend on the phase $\phi$, hence we infer the following useful relation
\begin{equation}\label{Elegant}
\sum_{i,j}\langle\hat{P}^\phi_i\hat{P}^\phi_j\rangle=\sum_{i,j}\langle\hat{X}^\phi_i\hat{X}^\phi_j\rangle=\nu^{-1} \langle \sum_j\hat{P}^\phi_j\rangle.
\end{equation}
On the other hand, the result \eqref{trigon2} leads to
\begin{equation}\label{CorrApprox}
\sum_{i,j}\langle\hat{X}^{\bar{\phi}}_i\hat{X}^{\bar{\phi}}_j\rangle\approx\sum_{i,j}\langle\hat{P}^\phi_i\hat{P}^\phi_j\rangle.
\end{equation}
Consequently, by putting together the relations (\ref{Elegant},\ref{Pfinal},\ref{Xfinal},\ref{CorrApprox}), we readily find the QFI.
\begin{align}
F_Q[\rho_{|\epsilon|}]&=\frac{\left(\frac{\partial \langle\hat{P}^\phi_{\rm sum}\rangle}{\partial |\epsilon|}\right)^2}{\Delta^2\langle\hat{P}^\phi_{\rm sum}\rangle}\approx \frac{2n_0N^2}{C_{\rm p}\kappa}, \\
F_Q[\rho_{\phi}]&=\frac{\left(\frac{\partial \langle\hat{X}^{\bar{\phi}}_{\rm sum}\rangle}{\partial \delta\phi}\right)^2}{\Delta^2\langle\hat{X}^{\bar{\phi}}_{\rm sum}\rangle}\approx \frac{2n_0N^2|\epsilon|^2}{C_{\rm p}\kappa}.
\end{align}
\bibliography{biblio}
\end{document}